\documentclass[aps,prb,twocolumn,superscriptaddress]{revtex4}

\usepackage{graphicx,color}

\input epsf.sty
\usepackage{graphicx}

\begin{document}

\preprint{\today}


\title{
High-energy magnetic excitations in overdoped La$_{2-x}$Sr$_{x}$CuO$_{4}$ studied by neutron and 
resonant inelastic X-ray scattering
}

\author{S. Wakimoto
}
\affiliation{ Quantum Beam Science Directorate, Japan Atomic Energy Agency,
   Tokai, Ibaraki 319-1195, Japan }

\author{K. Ishii}
\affiliation{ SPring-8, Japan Atomic Energy Agency,
   Hyogo 679-5148, Japan }

\author{H. Kimura}
\affiliation{ Institute of Multidisciplinary Research for Advanced Materials, 
   Tohoku University, Sendai 980-8577, Japan }

\author{M. Fujita}
\affiliation{ Institute for Materials Research, Tohoku University, Katahira,
   Sendai 980-8577, Japan }

\author{G. Dellea}
\affiliation{ CNR-SPIN and Dipartimento di Fisica, Politenico di Milano, Piazza Leonardo da Vinci32
   I-20133 Milano, Italy }

\author{K. Kummer}
\affiliation{European Synchrotron Radiation Facility, 6 rue Jules Horowitz,
   F-38043 Grenoble, France}

\author{L. Braicovich}
\affiliation{ CNR-SPIN and Dipartimento di Fisica, Politenico di Milano, Piazza Leonardo da Vinci32
   I-20133 Milano, Italy }

\author{G. Ghiringhelli}
\affiliation{ CNR-SPIN and Dipartimento di Fisica, Politenico di Milano, Piazza Leonardo da Vinci32
   I-20133 Milano, Italy }

\author{L. M. Debeer-Schmitt}
\affiliation{ Instrument and Source Division, Oak Ridge National Laboratory,
   Oak Ridge, Tennessee 37831, USA }

\author{G. E. Granroth}
\affiliation{ Quantum Condensed Matter Division, and Neutron Data Analysis and Visualization Division, Oak Ridge National Laboratory,
   Oak Ridge, Tennessee 37831, USA }

\date{\today}

\begin{abstract}

We have performed neutron inelastic scattering and resonant inelastic X-ray 
scattering (RIXS) at the Cu-$L_3$ edge to study high-energy magnetic excitations 
at energy transfers of more 
than 100 meV for overdoped La$_{2-x}$Sr$_{x}$CuO$_{4}$ with $x=0.25$ ($T_c=15$~K) and 
$x=0.30$ (non-superconducting) using identical single crystal samples for the two techniques.
From constant-energy slices of neutron scattering cross-sections, we have 
identified magnetic excitations up to $\sim 250$~meV for $x=0.25$.
Although the width in the momentum direction is large, the peak positions along the 
$(\pi, \pi)$ direction agree with the dispersion relation of the spin-wave in the 
non-doped La$_{2}$CuO$_{4}$ (LCO), which is consistent with the previous RIXS results of cuprate 
superconductors.
Using RIXS at the Cu-$L_3$ edge, we have measured the dispersion relations of the so-called 
paramagnon mode along both $(\pi, \pi)$ and $(\pi, 0)$ directions.
Although in both directions the neutron and RIXS data connect with each other 
and the paramagnon along $(\pi, 0)$ 
agrees well with the LCO spin-wave dispersion, the paramagnon in the $(\pi, \pi)$ direction 
probed by RIXS appears to be less dispersive and the 
excitation energy is lower than the spin-wave of LCO near $(\pi/2, \pi/2)$.
Thus, our results indicate consistency between neutron inelastic scattering and RIXS, and 
elucidate the entire magnetic excitation in the $(\pi, \pi)$ direction by the complementary 
use of two probes.  
The polarization dependence of the RIXS profiles indicates that appreciable charge excitations exist 
in the same energy range of magnetic excitations, reflecting the itinerant character of the overdoped 
sample.  
A possible anisotropy in the charge excitation intensity might explain the apparent differences in 
the paramagnon dispersion in the $(\pi, \pi)$ direction as detected by the X-ray scattering.

\end{abstract}


\maketitle

\section{Introduction}

High-temperature superconductivity in cuprates, such as La$_{2-x}$Sr$_{x}$CuO$_{4}$ (LSCO) and YBa$_{2}$Cu$_{3}$O$_{6+d}$ (YBCO), appears in the characteristic regime between the insulating antiferromagnetic and overdoped metallic regimes.  Considering that the magnetic fluctuations play an important role in the superconductivity of these compounds, describing the dynamical magnetic response in the superconducting regime is an important issue.

It is well understood that the magnetic excitations in undoped La$_{2}$CuO$_{4}$ (LCO) can be described by the spin-wave theory with nearest-neighbor, next-nearest-neighbor, and cycle-exchange terms.~\cite{Headings}
In contrast, neutron scattering experiments have revealed a characteristic magnetic excitation in the superconducting regime~\cite{Birgeneau_review,Fujita_review}: an incommensurate magnetic signal apparently disperses inward below a certain energy, $E_{cross}$, and then disperses outward above $E_{cross}$.~\cite{Tranquada_HG,Hayden_HG} This excitation, called an ``hour-glass'' excitation, is observed in a wide doping range of LSCO from non-superconducting $x=0.03$ (Ref.~\onlinecite{Matsuda_UD}) to slightly overdoped $x=0.16$.~\cite{Hayden_opt}  In the underdoped region, the low-energy incommensurability and $E_{cross}$ increase linearly with doping.~\cite{Matsuda_UD,Yamada_Yplot}  The low-energy incommensurate spin fluctuation appears up to the overdoped region; however, its cross section decreases linearly with superconducting transition temperature $T_c$ as the superconductivity decreases with overdoping.~\cite{Wakimoto_OD1,Wakimoto_OD2,Lipscombe_OD}  These facts suggest that the hole doping strongly affects the low-energy magnetic excitations and that they are closely related to the superconductivity.

An alternative to neutron inelastic scattering for the observation of single magnons is provided by the recently developed resonant inelastic X-ray scattering (RIXS) technique at the Cu-$L_3$ edge.~\cite{Braicovich_LCO,RIXS_review}   
In contrast to RIXS with the Cu-$K$ edge, which is more sensitive to charge excitations triggered by the core-hole potential in the 1$s$ orbital, RIXS with the Cu-$L_3$ edge can trigger a single magnon excitation by the spin-orbit coupling of 2$p$ orbitals.~\cite{L3RIXS_theory}  A systematic study of paramagnons by this technique using the hole-doped YBCO~\cite{LeTacon_natphys} and LSCO~\cite{Dean_2013} family compounds revealed that the dispersion relation along the $(\pi, 0)$ direction above 150~meV is nearly independent of the hole concentration in a wide doping range from the undoped to the non-superconducting overdoped samples.  This fact is in sharp contrast to the neutron results in the low-energy region.

\begin{figure*}
\includegraphics[width=17cm]{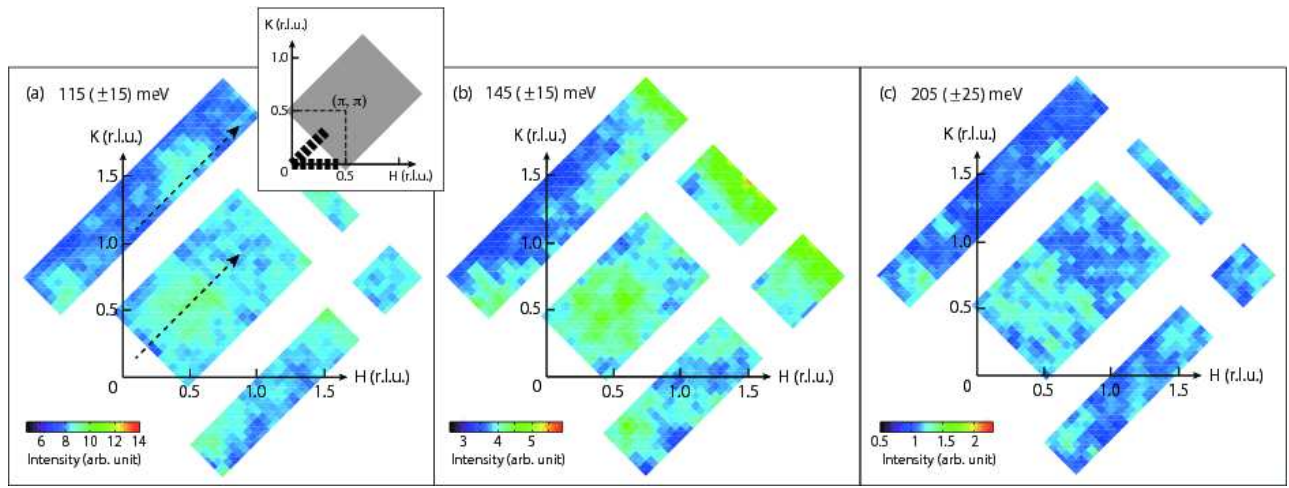}
\caption{(Color online) Contour maps of neutron scattering intensity in the $(H,K)$
zone at (a) $115(\pm15)$~meV, (b) $145(\pm15)$~meV, and (c) $205(\pm25)$~meV
energy transfer.  The antiferromagnetic zone center $(\pi, \pi)$ corresponds 
to $(0.5, 0.5)$ in this tetragonal notation.  Dashed arrows in (a) indicate 
trajectories of $(H,H)$ and $(H,1+H)$, on which intensity profiles are analyzed by fits.
The inset shows the scan area for neutron and RIXS.  The gray area indicates the 
coverage of detector banks of the neutron measurement that covers the antiferromagnetic 
zone center $(\pi, \pi)$.  Thick dashed lines indicate the trajectories of RIXS measurements 
by rotating the sample.
}
\end{figure*}

Because of the limited wave vector of incident photons at the Cu-$L_3$ edge and relatively relaxed energy resolution of 120~meV, compared with neutron inelastic scattering, RIXS at the Cu-$L_3$ edge is more appropriate for the measurement of magnon excitations above 100~meV dispersing from the $(0, 0)$ position, where the magnetic structure factor for neutron scattering is small.  In contrast, neutron scattering has a fine energy resolution, typically a few meV, which makes it difficult to observe high-energy magnetic excitation in doped samples, which is broad in energy, and consequently, neutron inelastic scattering is suitable for the observation of magnetic excitations below 150 meV, dispersing from the antiferromagnetic (AF) zone center $(\pi, \pi)$ for the doped samples.  Thus, RIXS and neutron inelastic scattering measurements are complementary to each other.

In this paper, we report high-energy magnetic excitations above 100 meV of overdoped LSCO studied by both neutron inelastic scattering and Cu-$L_3$ edge RIXS using identical crystals for the two techniques.  The overdoped sample was selected such that we can also test the doping independence of the paramagnon dispersion relation.  The neutron scattering results of LSCO $x=0.25$ indicate that the magnetic dispersion relation along the $(\pi, \pi)$ direction up to 250 meV reasonably agrees with the spin-wave dispersion relation of LCO, which is consistent with the doping-independence of magnetic dispersion reported using RIXS.  The Cu-$L_3$ edge RIXS measurements of the identical sample reveal that the dispersion along the $(\pi, 0)$ direction is consistent with the LCO spin-wave dispersion and previous RIXS results by Dean {\it et al.}~\cite{Dean_2013} using LSCO thin films.  In contrast, the paramagnon near $(\pi/2, \pi/2)$ appears to be less dispersive and the excitation energy near $(\pi/2, \pi/2)$ is lower than the spin-wave energy of LCO.

\section{Experimental procedure}

Single crystals of LSCO with $x=0.25$ and $0.30$ used for the neutron and Cu-$L_3$ RIXS measurements were grown using the traveling solvent floating zone method.  The growth and post-annealing conditions are the same as those described in Ref.~\onlinecite{Wakimoto_OD1}.  Single crystals of $x=0.25$ were used for the neutron measurements, and both $x=0.25$ and $0.30$ were used for the neutron inelastic scattering and RIXS.  The crystal structures of $x=0.25$ and $x=0.30$ are tetragonal with space group $I4/mmm$.  In the present study, we use the notation based on this tetragonal structure.  Therefore, the antiferromagnetic wave vector $(\pi, \pi)$ corresponds to $(0.5, 0.5)$.

Neutron scattering experiments were performed using the SEQUOIA chopper spectrometer~\cite{Granroth2006,Granroth2010} at the Spallation Neutron Source (SNS) at Oak Ridge National Laboratory (ORNL).  Five single crystals of $x=0.25$, with volumes of $\sim 1.4$~cm$^3$, were co-aligned using a neutron diffractometer installed at the CG-1B beam port of the High Flux Isotope Reactor (HFIR) at ORNL before the SEQUOIA measurements.  The set of co-aligned crystals was attached to the cold finger of a $^3$He closed cycle refrigerator and set on the spectrometer. 
An incident neutron Energy ($E_i$) of either $250$~meV or $350$~meV was selected by the coarse resolution fermi chopper spinning at 240Hz.  For the 250meV and 350meV settings, the T0 chopper was spun at 60Hz and 150 Hz, respectively.
For both conditions the $c$ axis of the crystal was oriented parallel to the incident beam.
These experimental conditions provides an energy resolution of 5 -- 20~meV, and an momentum resolution of approximately $0.03$~\AA$^{-1}$ along the $(\pi, \pi)$ direction for the energy transfer between 100 -- 250~meV.  

Cu-$L_3$ edge RIXS experiments were performed using the AXES spectrometer at the ID08 beam line of the European Synchrotron Radiation Facility (ESRF).  Crystals of $x=0.25$ and $0.30$ were cut into disk shapes with the $c$-axis normal to the disk surface.  The $x=0.25$ crystals were cut from the same crystal rod used for the neutron measurements.  The crystals were pasted on a Cu plate and attached to a refrigerator.  Either the $[1, 0]$ or $[1, 1]$ axis was set horizontal to enable the measurement of the paramagnon dispersion along the $(\pi, 0)$ or $(\pi, \pi)$ direction, respectively, by a horizontal rotation of the sample.  The incident photon polarization was set either horizontal ($\pi$-polarization) or vertical ($\sigma$-polarization).
The combined (monochromator for incident photons, spectrometer for scattering photons) energy resolution was $\sim 290$~meV.
Beam aperture from the sample to the detector is approximately 10 mRad, which leads to a momentum resolution of $0.004$~\AA$^{-1}$ with the photon energy at Cu-$L_3$ edge.
The scattering angle was 130$^{\circ}$.  Individual RIXS spectra obtained after 5 min of accumulation were summed for a total of 120 min at each $q$ value and polarization.  The energy spectra were measured every 5$^{\circ}$ of sample rotation, corresponding to less than 0.05 reciprocal lattice units (r.l.u.).

\begin{figure}
\includegraphics[width=8.5cm]{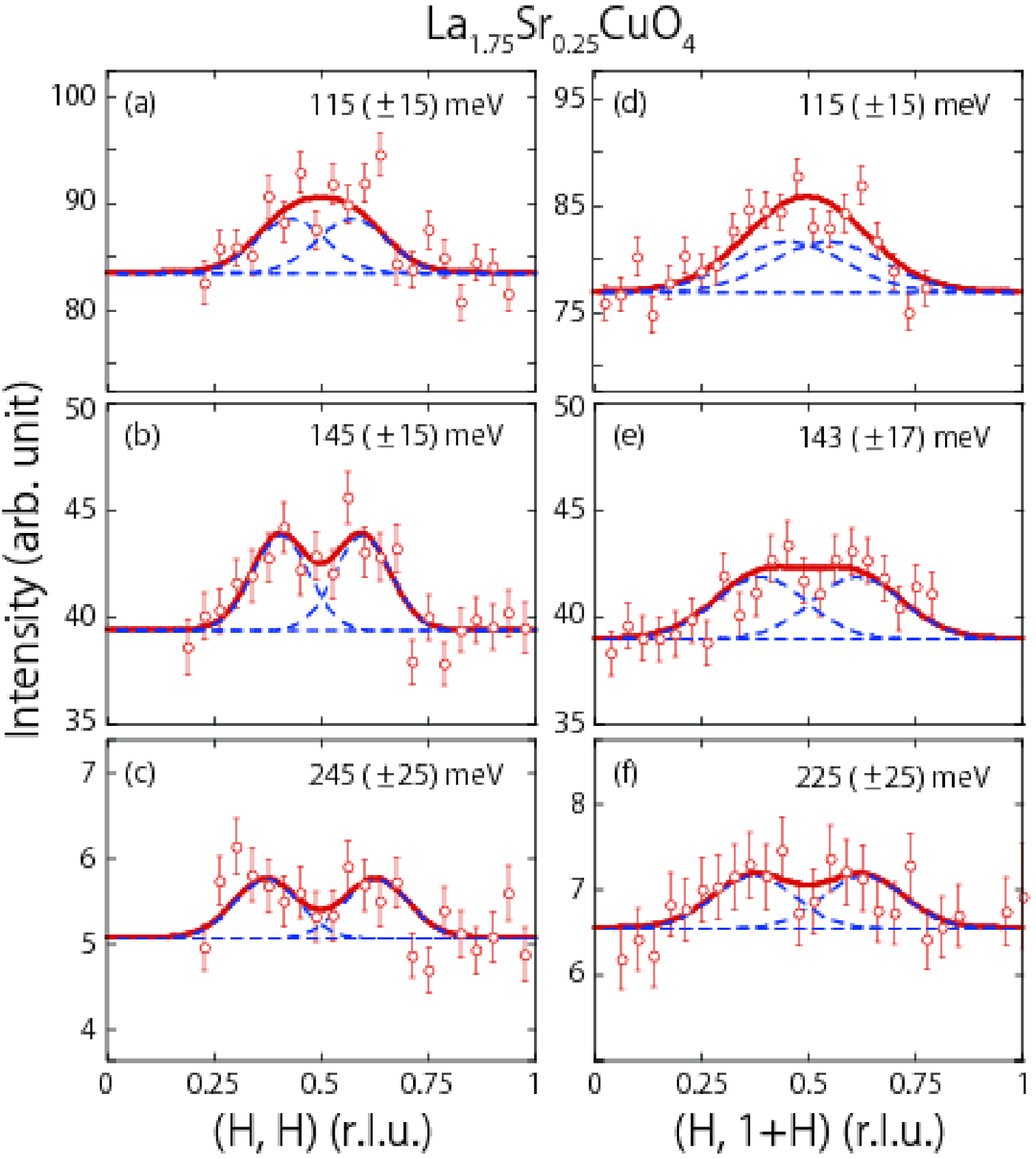}
\caption{(Color online) Neutron scattering profiles along trajectories $(H,H)$ and $(H,1+H)$
at selected energies.  Red lines are fits to a function containing two Lorentzians
symmetric to the AF zone center.  Each Lorentzian component is shown by a dashed line.
}
\end{figure}

\section{Neutron scattering results}

Neutron magnetic cross-sections below 100 meV of LSCO $x=0.25$ have already been reported in Ref.~\onlinecite{Wakimoto_OD2}.  Here, we focus on magnetic scattering in the energy range above 100~meV to compare the paramagnon dispersion measured by RIXS.  The magnetic excitation signal in the high-energy region becomes broad in energy upon hole doping.  Furthermore, the overall spectral weight in the overdoped region is smaller than that in the underdoped samples.  Therefore, we utilized a relatively poor energy resolution of 5 -- 20~meV, for these measurements.  Figure 1 presents contour maps of neutron cross-sections on the $(H, K)$-plane at energy transfers of $115$, $145$, and $205$~meV.  In Fig. 1 (a), a magnetic signal can be observed near $(0.5, 0.5)$, namely, the $(\pi, \pi)$ position.  This signal appears to become more dispersed from the AF zone center as the energy transfer increases.

\begin{figure}
\includegraphics[width=8.5cm]{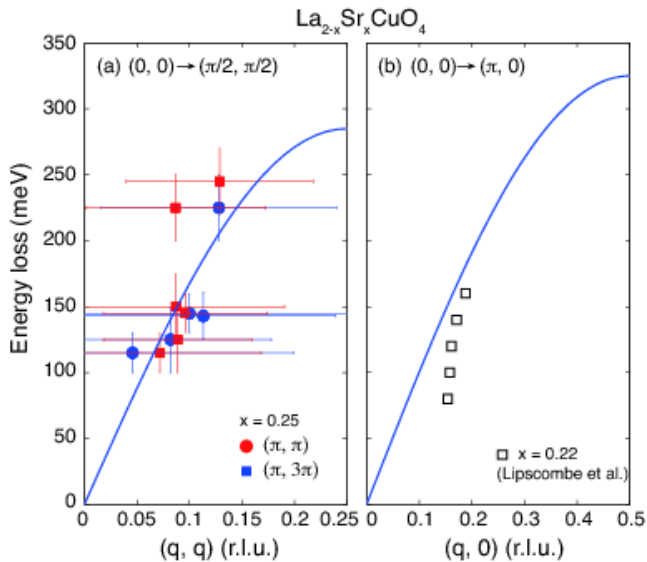}
\caption{(Color online) (a) Peak positions obtained by fits.  Vertical bars 
indicate the energy range of integration for the data analyses, and horizontal bars
indicate full widths at half maximum of Lorentzian peaks. (b) Magnetic peak positions of 
LSCO $x=0.22$ measured by neutron referred from Ref.\onlinecite{Lipscombe_OD}.  In both figures, 
solid lines are the spin-wave dispersion of non-doped LCO referred from Ref.\onlinecite{Headings}.
}
\end{figure}

To draw the magnetic dispersion, we analyzed intensity profiles along the trajectories of $(H, H)$ and $(H, 1+H)$ across the two AF zone centers, namely $(\pi, \pi)$ and $(\pi, 3\pi)$.  These trajectories are indicated by dashed arrows in Fig. 1 (a).~\cite{trajectory}  We fit the profiles by a two-Gaussian function, which is symmetric to the AF zone center.  We present select profiles with the results of the fits in Fig. 2 as a representative data set.  Figures 2 (a) -- (c) present profiles around $(\pi, \pi)$ at $115$, $145$, and $250$~meV, and Figs 2 (d) -- (f) present profiles around $(\pi, 3\pi)$ at $115$, $143$, and $225$~meV, respectively.  Here, it is more clearly demonstrated that the magnetic signals disperse outward as the energy increases consistently at both AF zones.  

The peak positions and full-width at half-maximum (FWHM) values obtained by the fits are summarized in Fig. 3 (a).  Here, the closed symbols represent the peak positions and the horizontal bars represent the FWHM values.  Because the analyzed scan trajectories are the $(H, H)$ directions, the peak positions correspond to the magnetic dispersion relation along $(\pi, \pi)$ from the magnetic zone center.  The solid curved line is the spin-wave dispersion along $(\pi, \pi)$ of LCO referred from Ref.~\onlinecite{Headings}.  The observed magnetic peak is broad in $q$ in this energy range; however, the peak positions roughly follow the spin-wave dispersion.  This finding is consistent with the doping independence of the paramagnon dispersion observed using RIXS for the $(\pi, 0)$ direction.

It is worth comparing the present data to the data of LSCO $x=0.22$ reported by Lipscombe {\it et al.}~\cite{Lipscombe_OD} in the same energy range ($> 100$~meV).  Our results indicate that the half-width at half-maximum (HWHM) is $\sim 0.21$~\AA$^{-1}$ for the profiles around $(\pi, \pi)$ and $\sim 0.30$~\AA$^{-1}$ for those around $(\pi, 3\pi)$.  At this stage, the reason for the difference in the different zones is unknown; however, the values are apparently consistent with that of LSCO $x=0.22$.  Lipscombe {\it et al.} reported that the width $\kappa$ suddenly increases to $0.3$~r.l.u. above $70$~meV.  Note that the functional form they used to analyze the data is different from ours and that their $\kappa$ is larger than our HWHM by a factor of $\sim 1.6$.  Thus, the width of LSCO $x=0.22$ corresponding to our HWHM is $\sim 0.31$~\AA$^{-1}$, which is in reasonable agreement with our values.

Next, we compare the dispersion relation.  In Ref.~\onlinecite{Lipscombe_OD}, the incommensurability $\delta$, which approximately corresponds to the peak position in our analyses, along the $(\pi, 0)$ direction is reported.  The open symbols in Fig.3 (b) show $\delta$ values of LSCO $x=0.22$ between $80$ and $160$~meV together with the spin-wave dispersion of LCO along $(\pi, 0)$~\cite{Headings}.  In contrast to the $(\pi, \pi)$ direction data, the peak positions along $(\pi, 0)$ of LSCO $x=0.22$ clearly deviate from the spin-wave dispersion relation, although it appears that the peak position approaches the spin-wave dispersion of LCO as the energy increases.  In the next section, we present the Cu-$L_3$ RIXS results and compare these findings with the neutron results.

\begin{figure}
\includegraphics[width=8.5cm]{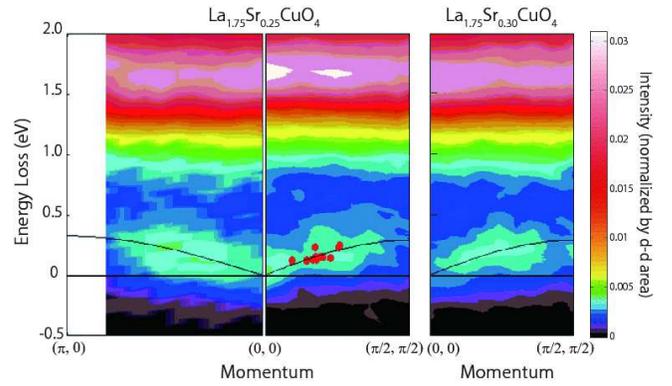}
\caption{(Color online) Contour maps of RIXS intensity for $x=0.25$ and $0.30$.  
The high intensity at $\sim1.8$~eV
is due to the $dd$ excitation.  Dispersive feature below 0.5~eV is the paramagnon.
Solid dispersive lines are the spin-wave dispersion of LCO.  Solid circles indicate the 
peak positions of neutron magnetic peak in Fig. 3. 
}
\end{figure}

\section{C\lowercase{u}-L$_{3}$ edge RIXS results}

Cu-$L_3$ RIXS profiles were measured using single crystals of $x=0.25$ and $0.30$ at several $q$-positions between $(0, 0)$ and $(\pi, 0)$ and between $(0, 0)$ and $(\pi, \pi)$.  Figure 4 presents contour maps of the RIXS intensity measured with the $\pi$-polarization configuration.  The intensity is normalized by the integrated intensity of the $dd$-excitation, which appears at $\sim 1.6$~eV as a dispersionless excitation.  The contour maps clearly indicate dispersive modes below $500$~meV for both the $(\pi, 0)$ and $(\pi, \pi)$ directions.  
In the same figure, the spin-wave dispersion of LCO is represented by solid lines, and the magnetic peak positions determined by neutron inelastic scattering along the $(\pi, \pi)$ direction are represented by circles.  Below, we compare these data in detail.

\begin{figure}
\includegraphics[width=8.5cm]{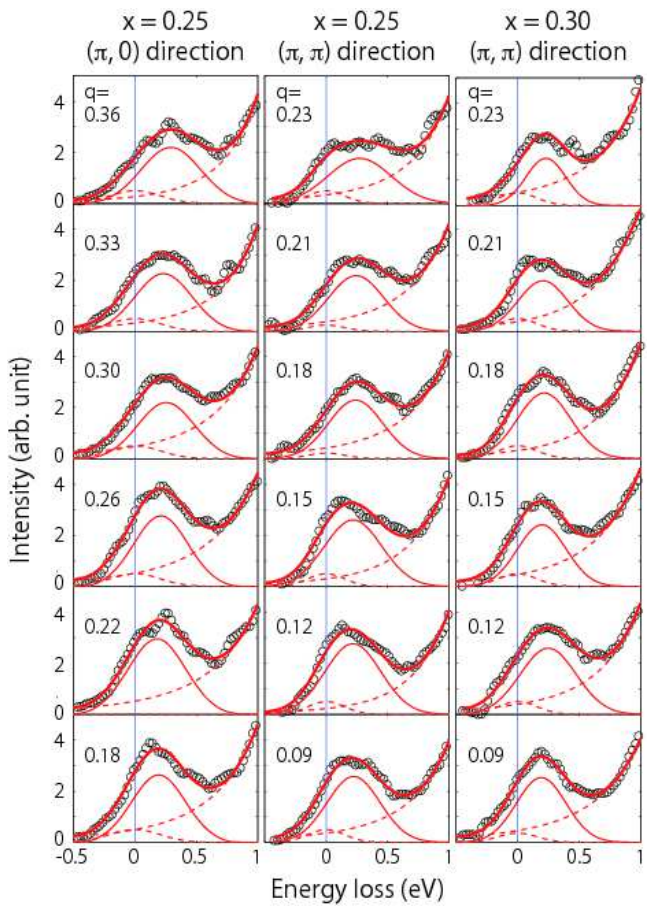}
\caption{(Color online) Cu-$L_3$ RIXS profiles.  Data were analyzed by fitting to a function
contains two Gaussians: one for magnon and the other for the $E=0$ components, and 
a squared Lorentzian describing 
the tail of the $dd$ excitation.  The $E=0$ component is assumed to have resolution width 
0.35~eV, while the magnon component is convoluted by the resolution. 
Thick solid lines are results of fits, thinner solid lines are 
magnon components, and dashed lines represent the $E=0$ and $dd$-tail components.
}
\end{figure}

To evaluate the energy of the paramagnon modes, we fit the RIXS profiles to a function containing a quasielastic peak (containing an elastic peak and phonon excitations), a paramagnon peak, and the tail of the $dd$-excitation.  
The quasielastic and paramagnon peaks are assumed to be Gaussian functions, and the tail of the $dd$-excitations is assumed to be a squared-Lorentzian tail.
The quasielastic component is resolution limited, whereas the magnon component is convoluted with the energy resolution of $350$~meV.  In a previous RIXS study of paramagnons with better energy resolution of $130$~meV, the RIXS profiles were analyzed with more peak components such as phonon and two-magnon components.  In the present study, the relatively broad energy resolution prevents these lower energy modes from being distinguished; however, the major paramagnon contribution should be evaluated by the above analyses.

The results of the fits are summarized in Fig. 5.  The magnon component is indicated by the solid curves, and the elastic component and tail of the $dd$-excitation are represented by dashed curves.  The magnon energies determined by the fits of the RIXS data are represented by the filled symbols in Fig. 6: circles ($x=0.25$), squares ($x=0.30$), and diamonds ($x=0.26$ referred from Ref.~\onlinecite{Dean_2013}).  It is observed in Fig. 6 (b) that the present data of $x=0.25$ along the $(\pi, 0)$ direction are consistent with the data of $x=0.26$ reported by Dean {\it et al.}~\cite{Dean_2013} using a thin-film sample.  This fact indicates that our analyses are valid, and importantly, the magnons are consistently observable both in bulk crystals and thin films.  
Figure 6 demonstrates that the agreement between the paramagnon energies, represented by solid symbols, and the spin-wave dispersion of LCO, represented by solid curves, is excellent for the $(\pi, 0)$ direction, whereas it is somewhat poor in the $(\pi, \pi)$ direction.  The RIXS data in the $(\pi, \pi)$ direction appear to be less dispersive and tend to be located at lower energies for $q$ larger than $(0.15, 0.15)$.   

\begin{figure}
\includegraphics[width=8.5cm]{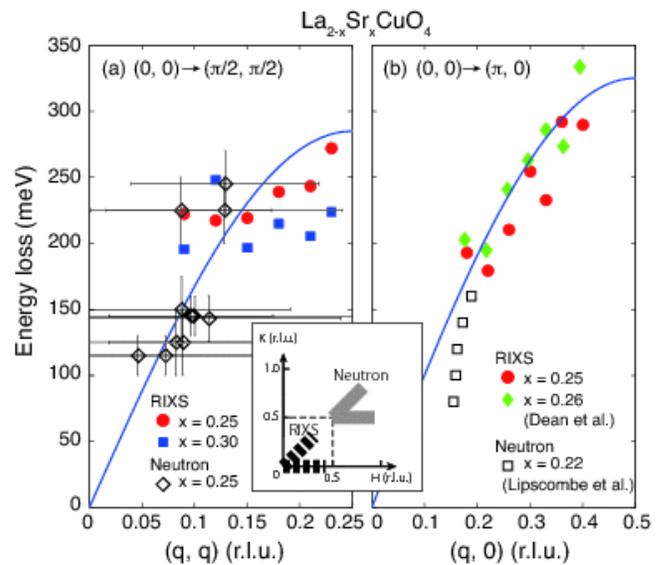}
\caption{(Color online) Paramagnon dispersions along (a) the $(\pi,\pi)$ direction for $x=0.25$ (circles) 
and $0.30$ (closed squares), and those along (b) the $(\pi,0)$ directions for $x=0.25$ (circles) 
and for $x=0.26$ thin film (closed diamonds) reported in Ref.\onlinecite{Dean_2013}.  
Neutron data in Fig. 3 are also shown by open squares.
Solid lines are the spin-wave dispersion of LCO.
The thick dashed and gray lines in the inset indicate q-trajectories of dispersions measured by RIXS 
and neutron, respectively.
}
\end{figure}

The data measured by neutron inelastic scattering are also shown in Fig. 6 as open symbols.  The data in Fig. 6 (a) are those of the present study and the data in Fig. 6 (b) are those of $x=0.22$ by Lipscombe {\it et al.}~\cite{Lipscombe_OD}  It should be noted that the RIXS data show the dispersion relation from the $(0, 0)$ position, whereas the neutron data from the AF zone center $(\pi, \pi)$ are shown in the inset of Fig. 6.  However, these data sets should be identical as long as the antiferromagnetic correlation exists.  
In both directions, the neutron and RIXS data connect with each other, thus indicating the consistency between these two probes and confirming that the spin excitation can be consistently observed by both probes.

\section{Discussion}

We have demonstrated that the magnetic excitations of overdoped LSCO single crystals measured by neutron and RIXS are qualitatively consistent with each other and that the high-energy magnetic dispersion in the $(\pi, 0)$ direction follows the spin-wave dispersion relation of LCO in the energy range above 150 meV.  
This behavior is consistent with the doping independence of magnon dispersion observed by RIXS for hole-doped cuprate thin films.\cite{LeTacon_natphys,Dean_2013,LeTacon_PRB}  
The magnetic dispersion along $(\pi, \pi)$ agrees less with the spin-wave dispersion; however, the overall energy scales roughly agree with the spin-wave dispersion.
The present results are in contrast with the excitations in electron-doped systems.  
By combining the neutron inelastic scattering and RIXS results, Ishii {\it et al.}\cite{Ishii_2014} and Lee {\it et al.}\cite{Lee_natphys} have reported that the magnetic excitation near the AF zone center becomes steeper as doping increases and the overall excitation energy shifts to higher energies with respect to the undoped system.  In contrast, our RIXS data for the $(\pi, \pi)$ direction suggest that the excitation energy near $(\pi/2, \pi/2)$ might be lower than the spin-wave energy.  Such asymmetry between hole- and electron-doped systems is consistent with the numerical calculation of the Hubbard model by Jia {\it et al.}\cite{Jia_2014}

In addition to the qualitative consistency with the theoretical calculations, our RIXS data along the $(\pi, \pi)$ direction in Fig. 6 (a) agree less with the spin-wave dispersion than those along the $(\pi, 0)$ direction.  Previous RIXS studies of cuprates using thin films mostly focused on the $(\pi, 0)$ direction and are consistent with the spin-wave dispersion.  
Recently, Guarise {\it et al.}\cite{Guarise_natcomm} and Dean {\it et al.}\cite{Dean_Bi} reported that the RIXS profiles of Bi-based cuprates along $(\pi, \pi)$ exhibit anomalous softening or dispersionless broad excitation.  The present RIXS data along $(\pi, \pi)$ also appear to be less dispersive than those along the $(\pi, 0)$ direction.  
In Fig. 7, we present a comparison of the RIXS spectra with $\pi$- and $\sigma$-polarization configurations at the position where 
the sample is rotated by 45$^{\circ}$ from the specular position.  The data shows appreciable RIXS intensities of the $\sigma$-polarization, which is even larger than the intensities of the $\pi$-configuration.
Theoretical calculations based on a single Cu$^{2+}$ ion where the valence band has the $x^2 - y^2$ orbital symmetry indicate that in our experimental geometry, the RIXS cross-sections with and without spin-flip processes dominate the $\pi$ and $\sigma$ incident polarization configurations, respectively.\cite{L3RIXS_theory,Moretti_11,Braicovich_prb10} 
Therefore, the results in Fig. 7 indicate that appreciable charge excitation exists in the same energy range of magnetic excitation.  These charge excitations may affect the magnetic excitations, possibly in different manners between the $(\pi, \pi)$ and $(\pi, 0)$ directions, causing the observed difference in the magnetic excitations.
To address this question, more precise RIXS measurements with finer energy resolution and polarization analyses are necessary to distinguish the magnetic and charge excitations.

\begin{figure}
\includegraphics[width=8.5cm]{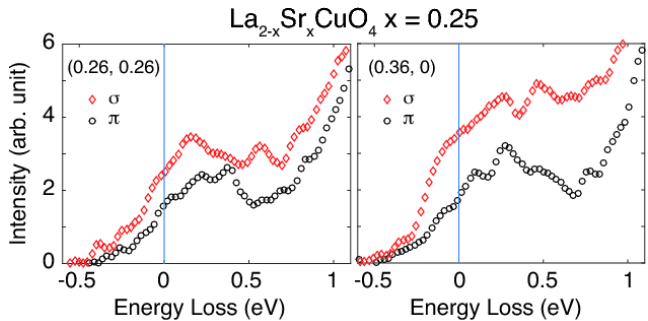}
\caption{(Color online) RIXS profiles measured with $\pi$ and $\sigma$ configurations at $(0.26, 0.26)$ and $(0.36, 0)$ corresponding to a 45-degree rotation of the sample from the specular orientation. 
}
\end{figure}

Remarkably, Fig. 7 shows that the RIXS intensity of the $\sigma$-configuration is always higher than that of the $\pi$-configuration.  This feature qualitatively resembles the $\pi$ and $\sigma$-configuration data of electron-doped Nd$_{2-x}$Ce$_{x}$CuO$_{4}$ with $x=0.15$ and $0.18$ reported in Ref.~\onlinecite{Ishii_2014}, but not those of the optimally hole-doped system in Ref.~\onlinecite{LeTacon_natphys}.  As discussed in Ref.~\onlinecite{Ishii_2014}, such a balance of the $\pi$ and $\sigma$-configuration spectra suggests that the spin and charge excitations are mixed because of itinerant character.  This finding suggests that in the electron-doped system, the itinerant character becomes stronger even at the optimally doped level and the magnetic excitation changes concomitantly, whereas in the hole-doped system, the itinerant character slowly grows with doping, and the high-energy magnetic excitation hardly changes with doping.  In contrast, our data indicates that the overdoped $x=0.25$ has itinerant character, but the magnetic excitation still remains mostly at the spin-wave dispersion of LCO.  This fact suggests that the asymmetry of the magnetic excitation between electron and hole-doped systems is not simply due to the stronger itinerant character of the electron-doped system.

\section{Summary}

We have performed neutron and Cu-$L_3$ edge RIXS measurements of overdoped La$_{2-x}$Sr$_x$CuO$_4$ using identical single crystals.  
The combination of neutron and RIXS indicates that both data sets are consistent with each other and that the overall high-energy magnetic excitation agrees with the spin-wave dispersion relation of the parent compound La$_2$CuO$_4$ particularly in the $(\pi, 0)$ direction, which is consistent with the previous RIXS studies using thin-films.  
We also draw the magnetic excitation above 100 meV in the $(\pi, \pi)$ direction by the complementary use of neutron and RIXS.  The magnetic excitation for the $q$ values smaller than $(0.15, 0.15)$~(r.l.u.) measured by neutron inelastic scattering follows the LCO spin-wave dispersion, whereas that for the $q$ values larger than $(0.15, 0.15)$~(r.l.u.) measured by RIXS is apparently less dispersive, and the excitation energy near $(\pi/2, \pi/2)$ is smaller than the LCO spin-wave excitation energy.  
The polarization dependence of the RIXS spectra indicates that appreciable charge excitations exist in the same energy range of magnetic excitations, which may affect the magnetic excitation.

\begin{acknowledgments}

Authors thank R. Kajimoto, M. Matsuda, M. Matsuura, K. Nakajima, T, Tohyama and K. Yamada for invaluable discussion. 
Authors also acknowledge L. J. Santodonate and M. Matsuda for their assistance in using CG1B at ORNL.
This work is supported by Grant-In-Aid for Scientific Research (C) No. 25390132 and No. 35400333, and by Grant-In-Aid for Scientific Research (B) No. 24340064.
Part of the research conducted at ORNL's High Flux Isotope Reactor and Spallation Neutron Source was sponsored by the Scientific User Facilities Division, Office of Basic Energy Sciences, US Department of Energy.
The RIXS experiments were performed using the AXES instrument at ID08 at the European Synchrotron Radiation Facility (ESRF).  We acknowledge ESRF for provision of synchrotron radiation facilities and we would like to thank N. B. Brookes for the support at ID08. 

\end{acknowledgments}


\begin{references}
\bibitem{trajectory}
We analyzed the data only on the trajectories along $(\pi, \pi)$ since the data along $(\pi, 0)$ was not able to be analyzed unambiguously due to the detector gap.
\end{references}

\end{document}